\documentclass[prb,aps,showpacs,showkeys,preprint]{revtex4}
\usepackage{graphicx}
\usepackage{amsmath}
\usepackage{amssymb}
\usepackage{amsthm}
\usepackage{booktabs}
\usepackage{stmaryrd}
\usepackage{longtable}
\usepackage[figuresright]{rotating}
\usepackage{multirow}
\usepackage{makecell}
\usepackage{dcolumn}
\usepackage{anysize}
\usepackage{color}
\marginsize{1.5cm}{1.5cm}{1.5cm}{1.5cm}
\newcolumntype{d}[1]{D{.}{.}{#1}}
\begin{document}
\title{Crystal field parameters with Wannier functions: application to rare earth aluminates}
\author{ P. Nov\'ak, K. Kn\'i\v{z}ek, and J. Kune\v{s}} %\email{novakp@fzu.cz}
\affiliation{Institute of Physics of ASCR, Cukrovarnick\'a 10, 162 53 Prague 6, Czech Republic}
\date{\today}
\begin{abstract}
A method to calculate the crystal field parameters is proposed and applied to trivalent rare earth
impurities in yttrium aluminate and to Tb$^{3+}$ ion in TbAlO$_3$. To determine crystal field
parameters local Hamiltonian  expressed in the basis of Wannier functions is expanded in a series
of spherical tensor operators. Wannier functions are obtained by transforming the Bloch functions
calculated using the density functional theory based program. The results show that the crystal
field is continuously decreasing as the number of $4f$ electrons increases and that the
hybridization of $4f$ states with the  states of oxygen ligands is important. The method contains
a single adjustable parameter characterizing the $4f$--ligand charge transfer.
Theory is confronted with experiment for Nd$^{3+}$ and Er$^{3+}$ ions in YAlO$_3$ matrix and for
Tb$^{3+}$ ion in TbAlO$_3$ and a good agreement within a few meV is found.

\end{abstract}

\maketitle
\section{Introduction}
{\it Ab initio} calculations of the properties of molecules and solids have become
a common tool of solid states physics and quantum chemistry.
Nevertheless open problems remain, the description of
the $4f$ states of rare earth (R) elements being one of them.
Above the Kondo temperature the physics of the $4f$ electrons is described
by an effective atomic Hamiltonian. As the Kondo temperatures (with the exception
of Ce or Yb compounds) of most rare earth impurities are well below
the experimental range of interest the effective atomic Hamiltonian
provides an important tool to study the $4f$ physics.
Parameters of the Hamiltonian may be fitted to experimental data or estimated
using semiempirical or {\it ab initio} methods.

The original motivation of this work was to explain the magnetic properties of rare earth
cobaltites RCoO$_3$ (R = rare earth). For these compounds few experimental data are available,
certainly not sufficient to estimate the crystal field parameters (CFP). Hence a necessity to
calculate CFP emerged. To this end we have developed a method described in this paper. To check
its reliability it was applied to orthorhombic rare earth aluminates, possessing the same crystal
structure as RCoO$_3$. Importantly, numerous experimental data exist for R$_x$Y$_{1-x}$AlO$_3$,
which are widely used in lasers, scintillator and optical recording media. In several cases
complete sets of CFP were deduced (see e.g. \cite{duan,gruber}).

The effective atomic Hamiltonian consists of the free ion interaction part $\hat{H}_A$ and the
one-particle crystal field term $\hat{H}_{CF}$
 \begin{equation}
\label{eq:h}
 \hat{H}=\hat{H}_A + \hat{H}_{CF}.
\end{equation}
The rotationally invariant free ion Hamiltonian
is only weakly material dependent.
It contains the energy in a central field (a trivial constant when
restricted to the $4f$ shell as usual),
electron-electron interaction and the spin-orbit coupling.
Details of $\hat{H}_A$ can be found in Ref. \onlinecite{hufner}. Carnall {\it et al.}
\cite{carnall} determined the parameters of $\hat{H}_A$ for all R$^{3+}$ ions
in LaF$_{3}$ by carefully fitting the optical absorption spectra.
The parameters for Nd$^{3+}$ and Er$^{3+}$ ions in
YAlO$_3$ \cite{duan} and for Tb$^{3+}$ ion in TbAlO$_3$ \cite{gruber} were
determined analogously.

Construction of the material specific $\hat{H}_{CF}$ presents a formidable theoretical problem.
The one-particle crystal field Hamiltonian can be written as \cite{wybourne}
\begin{equation}
 \hat{H}_{CF} = \sum_{k=0}^{k_{max}}\sum_{q=-k}^k B_{q}^{(k)} \hat{C}_{q}^{(k)},
\label{eq:hcf}
\end{equation}
where $ \hat{C}_{q}^{(k)}$ is a spherical tensor operator of rank $k$ acting on  electrons in the
$4f$ shell. $B_{q}^{(k)}$ are the crystal field parameters. For the $f$ electrons $k_{max}$ is equal to six,
providing that cross terms of $\hat{H}_{CF}$ between states of different angular momenta are neglected. Hermiticity of $\hat{H}_{CF}$ requires
that $B_{-q}^{k} = (-1)^q B_{q}^{k*}$.

The number of non-zero $B_{q}^{(k)}$ depends on the site symmetry.
For a low site symmetry this number may be large and it is not  {\it a priori} possible to predict,
which of the CFP are important. When analyzing experimental data the CFP are usually determined
by the least squares fit. This is often an ambiguous procedure with several equally likely solutions.
In magnetic or superconducting compounds the number of experimental data is
usually insufficient to determine CFP, yet the magnetic properties reflect the crystal field sensitively. There were therefore
numerous attempts to estimate CFP theoretically and the effort is continuing (see Ref. \onlinecite{novak3} for recent survey).

In the present paper we use Wannier functions to construct the crystal field Hamiltonian.
The method is applied to  R$^{3+}$ ions in orthorhombic aluminates. A detailed comparison with experimental data,
presented in section \ref{sec:comparison},
shows that the method is capable of an accurate prediction of the $4f$ crystal field parameters.

\section{Crystal field states of R$^{3+}$ ions in orthorhombic aluminates}
\label{sec:cf}
The  orthorhombic rare earth aluminates have  distorted perovskite structure belonging to the $D^{16}_{2h}$ space group.
The unit cell of RAlO$_3$ contains four formula units. The R sites with the point group $C_s$ are surrounded by twelve oxygen atoms.
Choosing the quantization axis along the orthorhombic $c$ axis
the horizontal symmetry plane causes all $B_{q}^{(k)}$ with odd $q$ to be zero,
leaving three real $B_{0}^{(k)}$ parameters ($k$= 2, 4, 6) and six independent complex parameters $B_{q}^{(k)}$ ($k$= 2, 4, 6; $q$ = 2, 4,
 6; $q \leqq k$). There are thus fifteen numbers to be determined.
The spectrum of Hamiltonian (eq. \ref{eq:h}) does not depend on its orientation with the respect to
the crystallographic axes. Therefore the imaginary part of a selected CFP can be eliminated by a
specific rotation around the quantization axis~\cite{rudowicz} and only fourteen independent CFP
are needed to describe the experimental multiplet structures while the angle of rotation remains
undetermined. Conventionally the $B_{2}^{(2)}$ parameter is set to be real. In the presented
method all fifteen $B_{q}^{(k)}$ are determined. The rotation is only invoked in section
\ref{sec:discussion} when comparing the calculated CFP with those obtained using the least squares
fit by Duan {\it et al.} \cite{duan} and Gruber {\it et al.} \cite{gruber}.

On one-particle level the $C_s$ crystal field splits the seven $4f$ states $|l,m\rangle$ ($l$=3, $m=\pm 3, \pm 2, \pm 1, 0)$
into seven orbital singlets, which are of two different types.
Four of these singlets are formed by $m= \pm 3$ and $m= \pm 1$ states, in the remaining three
singlets $m= \pm 2$ orbitals are mixed with the $m$ =0 state.
In our analysis instead of the $|l,m\rangle $ states the basis of real orbitals is used
\begin{alignat}{2}
\label{eq:psi_real}
\nonumber |\varphi_1\rangle & = \frac{i}{\surd 2}(|3,-3\rangle + |3,3\rangle)\sim  y(3x^2-y^2)\\
\nonumber |\varphi_2\rangle & = \frac{1}{\surd 2}(|3,-3\rangle - |3,3\rangle)\sim x(x^2-3y^2) \\
|\varphi_3\rangle & = \frac{i}{\surd 2}(|3,-1\rangle + |3,1\rangle) \sim yz^2 \\
\nonumber |\varphi_4\rangle & = \frac{1}{\surd 2}(|3,-1\rangle - |3,1\rangle) \sim xz^2 \\
\nonumber |\varphi_5\rangle & = \frac{1}{\surd 2}(|3,-2\rangle + |3,2\rangle)\sim z(x^2-3y^2) \\
\nonumber |\varphi_6\rangle & = \frac{i}{\surd 2}(|3,-2\rangle - |3,2\rangle)\sim xyz \\
\nonumber |\varphi_7\rangle & = |3,0\rangle \sim z^3
\end{alignat}

In terms of real orbitals the wave functions of the seven singlets may be written as
\begin{equation}
\label{eq:psi}
 \psi_i = \sum_{j=1}^4 c_{j,i}\; |\varphi_j\rangle ;\;\; i=1,2,3,4 ;\;\;\;
 \psi_i = \sum_{j=5}^7 c_{j,i}\; |\varphi_j\rangle ;\;\; i=5,6,7.
\end{equation}

\section{Calculation of electronic structure and description of method}
\label{sec:method}
The computational procedure consists of two steps.
The {\it initial step} of our analysis is the standard self-consistent solution
of the Kohn-Sham equations of the density functional theory. Here we use the augmented plane waves
+ local orbital method implemented in the WIEN2k program \cite{wien}. For the exchange-correlation
functional the generalized-gradient approximation form \cite{perdew} was adopted. The experimental
orthorhombic lattice parameters of YAlO$_3$ \cite{diehl}  were used for Y$_{1-x}$R$_x$AlO$_3$ and
TbAlO$_3$ \cite{tb_a_b_c} respectively, while the atomic positions within the unit cell were
optimized for each system. The typical concentration $x$ of R ions in Y$_{1-x}$R$_x$AlO$_3$ used
as laser materials varies between 0.01 and 0.03. In our calculations the unit cell contained 120
atoms (RY$_{23}$Al$_{24}$O$_{72}$, corresponding to $x$=0.0435) retaining the orthorhombic
symmetry. The eigenvalue problem was solved in four points of the irreducible Brillouin zone and
the number of basis functions was $\sim$ 9200 (corresponding to parameter $RK_{max}$=6.13). The
calculations were non-spin-polarized and the $4f$ electrons were treated as {\it core} electrons,
%KK electrons. => electrons,
which contribute to the spherical component of the density only. As a consequence, the potential
on the R site does not contain any non-spherical components arising from the on-site $4f$ states.
This is vital for determination of the crystal field parameters, as otherwise the non-physical
interaction of the $4f$ states with the non-spherical potential they themselves create
(self-interaction) would dominate CFP. Note that the core treatment of the $4f$ states is specific
for the augmented plane waves basis and other methods require different means to eliminate the
non-spherical part of the $4f$ self-interaction.

In the {\it second step} the effective crystal-field Hamiltonian for the $4f$ electrons is constructed
from ingredients involving the shape of the $4f$ orbitals, the effective potential and
hybridization with the ligand orbitals. To this end the $4f$(R) orbitals are included in the valence basis set.
Before the eigenvalue problem with the potential from the initial step
is solved the relative energy of the $4f$ and ligand states is
modified by means of an orbitally dependent potential. This correction mimics the effect of electron-electron
interaction within the $4f$ shell and we justify it as follows.

While the inter-atomic hopping parameters, determined largely by the orbital shapes and atomic distances,
are relatively insensitive to the local Coulomb interaction, the energy separation of the $4f$ and ligand states is problematic.
This is not surprising as we are trying to represent the physics of extremely correlated systems
(the $4f$ shell) by an effective single particle scheme. There is no such universal
representation. Instead, one can develop effective representations for specific quantities. For
example, LDA+U provides such an effective model for one-particle photoemission and inverse
photoemission spectra, i.e. transitions involving addition or removal of an electron. Here, we are
interested in crystal-field optical excitations and thermodynamic properties, involving effects in
which the number of electrons does not change. Therefore LDA+U is not applicable. The key quantity
that controls the impact of the $4f$--ligand hybridization on the multiplet structure of the rare
earth is the charge-transfer energy, the cost of moving an electron from the ligand to the $4f$
shell. The difference of the Kohn-Sham energies $\epsilon_f-\epsilon_p$ gives a poor estimate of
the charge-transfer energy in a strongly interacting system. Therefore we introduce a correction
$\Delta$, which amounts to a downward shift of the oxygen $s$ and $p$ levels. The purpose of the
correction is to modify the difference $\epsilon_f-\epsilon_p$ so that it approximates the actual
charge transfer energy of the real material. We present more detailed reasoning in the Appendix.
We eliminate the minor charge transfers (and corresponding level corrections) to other than oxygen
orbitals by removing those from our basis set. The associated uncertainty of CFP is discussed in
the Results section.

Once the eigenvalue problem is set up the remaining analysis reduces to algebraic manipulations.
We proceed by transforming the Bloch states from the $4f$ energy window to Wannier functions using
the wien2wannier interface \cite{kunes} followed by standard application of the wannier90 software
\cite{wannier}. Wannier90 provides the seven by seven matrix $\hat{H}_{4f}$ of the Hamiltonian
between the $4f$ Wannier functions centered on the same atom, the traceless part of which is the
desired crystal field Hamiltonian $\hat{H}_{CF}$ (\ref{eq:hcf})
\begin{equation}
 \hat{H}_{4f} = E_{avg}\hat{I} + \hat{H}_{CF} =  E_{avg}\hat{I} + \sum_{k,q} B_{q}^{(k)} \hat{C}_{q}^{(k)},
\end{equation}
where
\begin{equation}
E_{avg}=\mathrm{Tr}(\hat{H}_{4f}/7).
\end{equation}
To get the CFP in the standard form we transform $\hat{H}_{4f}$ into the basis of spherical harmonics
and expand it as a 49-dimensional vector in the basis of spherical operators.

Summarized the computational procedure reads: (i) non-spin-polarized self-consistent WIEN2k calculation with $4f$
in core, (ii) solution of the eigenvalue problem with $4f$ in the valence window shifted with respect
to the ligand states, (iii) construction of the Wannier functions for the energy window of the $4f$ states
and extracting the on-site Hamiltonian, (iv) expansion of the on-site Hamiltonian in terms of the spherical operators
to obtain the CFP. We note that concept of the local Hamiltonian is also in the heart of CFP calculation method
proposed recently by Hu {\it et al.} \cite{hu}.

The method was first tested by determining the crystal field parameters of Pr$^{4+}$ ion in PrO$_2$ \cite{novak3}. The results were similar to those
reported by us earlier \cite{novak2}, albeit without the uncertainty in decomposition of the density of states peaks.
The symmetry of Pr$^{4+}$ site in PrO$_2$ is cubic and two real CFP are sufficient to characterize the crystal field.
The $4f$ states are split in two triplets and singlet and it is straightforward to determine CFP using the energy differences
only. This, however, does not hold when the symmetry is lower: in orthorhombic aluminates the $4f$ states are split in
seven singlets (\ref{eq:psi}) and six energy differences are certainly insufficient to determine fifteen CFP. From the
density of states projected on the $4f$ orbitals the orbital composition of the singlets, characterized by absolute values $|c_{ij}|$ in
 (\ref{eq:psi}) may also be extracted, increasing the number of data obtained from DOS to 25. Our attempt to determine
CFP using these data by the least squares led to similar ambiguity as encountered when analyzing the experimental
results - several almost equivalent solutions were found. In contrast, the method based on decomposition of the local Hamiltonian
gives the CFP set unambiguously.

\section{Results}
\label{sec:results}
\begin{figure}
\includegraphics[width=12cm]{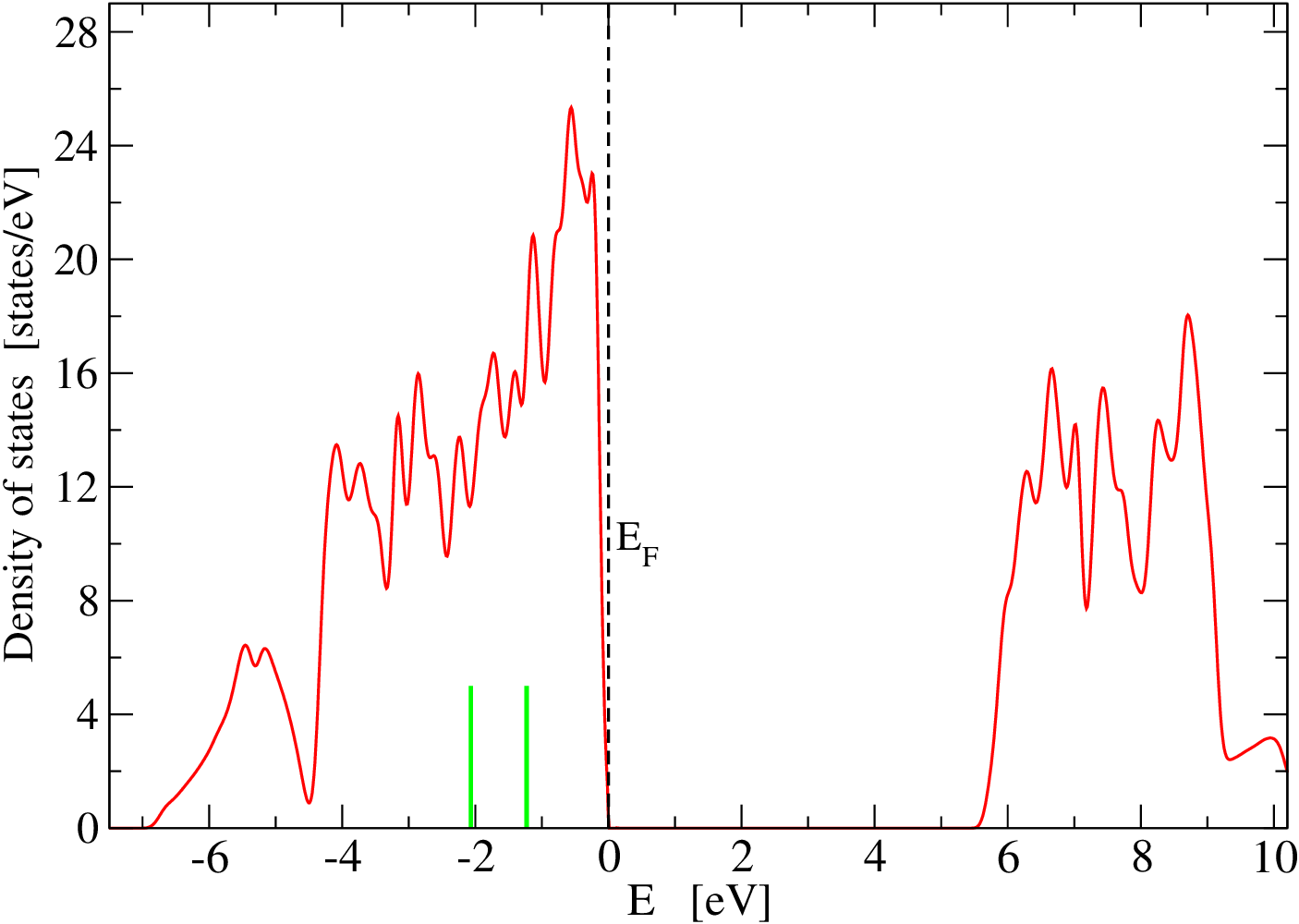}
\caption{(Color online) Total density of states of TbY$_{23}$Al$_{24}$O$_{72}$ with $4f$ states included in the core. Fermi energy is at zero,
two vertical lines indicate the positions of $4f$ core levels, split by the spin-orbit coupling.}
\label{fig:totdos}
\end{figure}

The calculations with the $4f$(R) states in the core result
in insulating band structures with the gaps ranging from
5.63 eV for R=Ce to 5.76 eV for R=Yb in RY$_{23}$Al$_{24}$O$_{72}$
and amounting to 5.04 eV in TbAlO$_3$.
The valence band is dominated by oxygen $2p$ states,
while the bottom of the conduction band is formed mainly by $5d$(R) and $4d$(Y) orbitals.
An example of the density of states from such calculation is shown
in Fig. \ref{fig:totdos}.

\begin{figure}
\includegraphics[width=12cm]{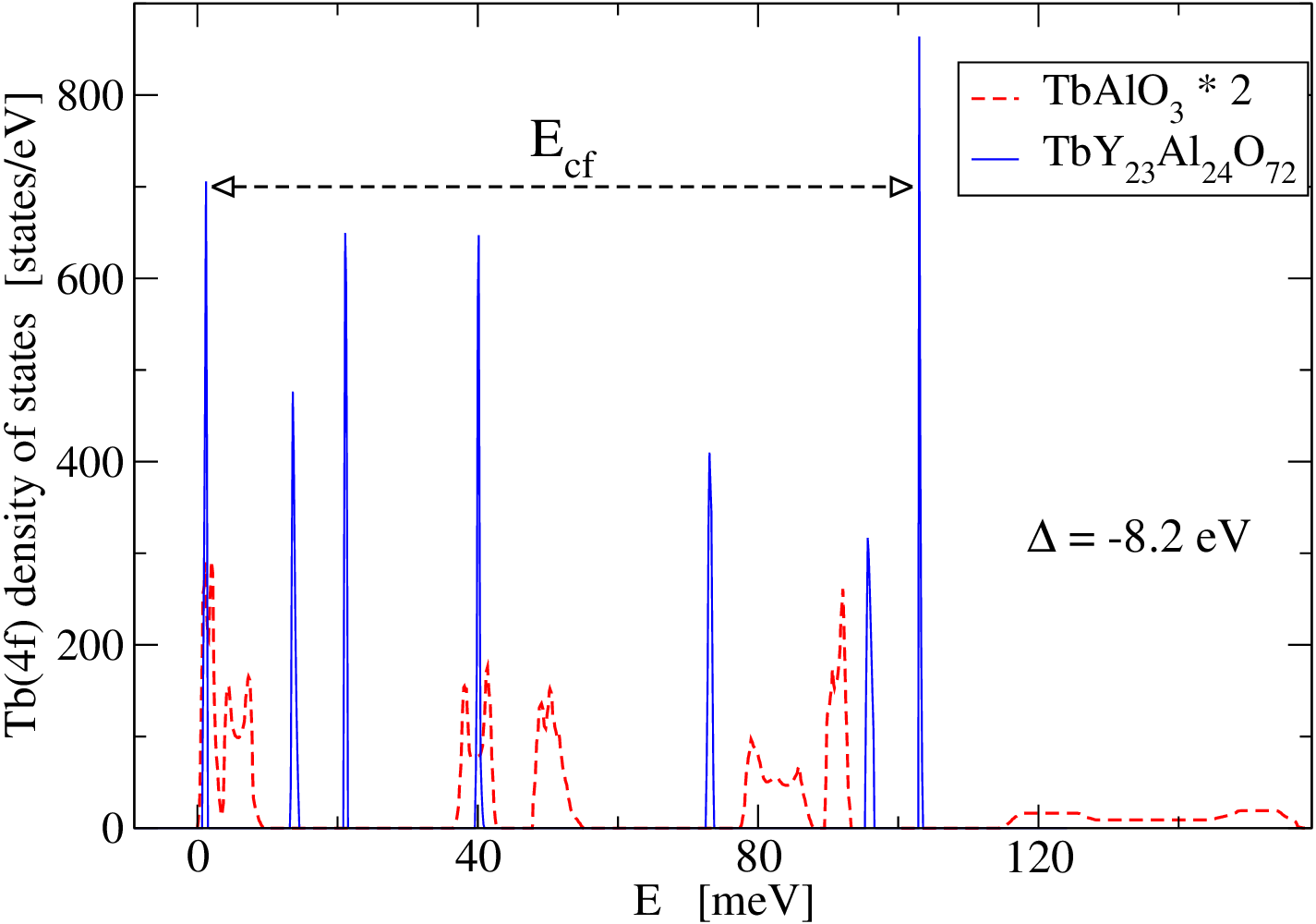}
\caption{(Color online) TbAlO$_3$ and Tb:YAlO$_3$. Density of states projected on the $4f$ subspace. The shift $\Delta$ equals to -8.2 eV.
$E_{cf}$ is the energy difference between the highest and lowest $4f$ singlet states and it is used to characterize the strength of the crystal field.}
\label{fig:tbdos}
\end{figure}

\begin{figure}
\includegraphics[width=12cm]{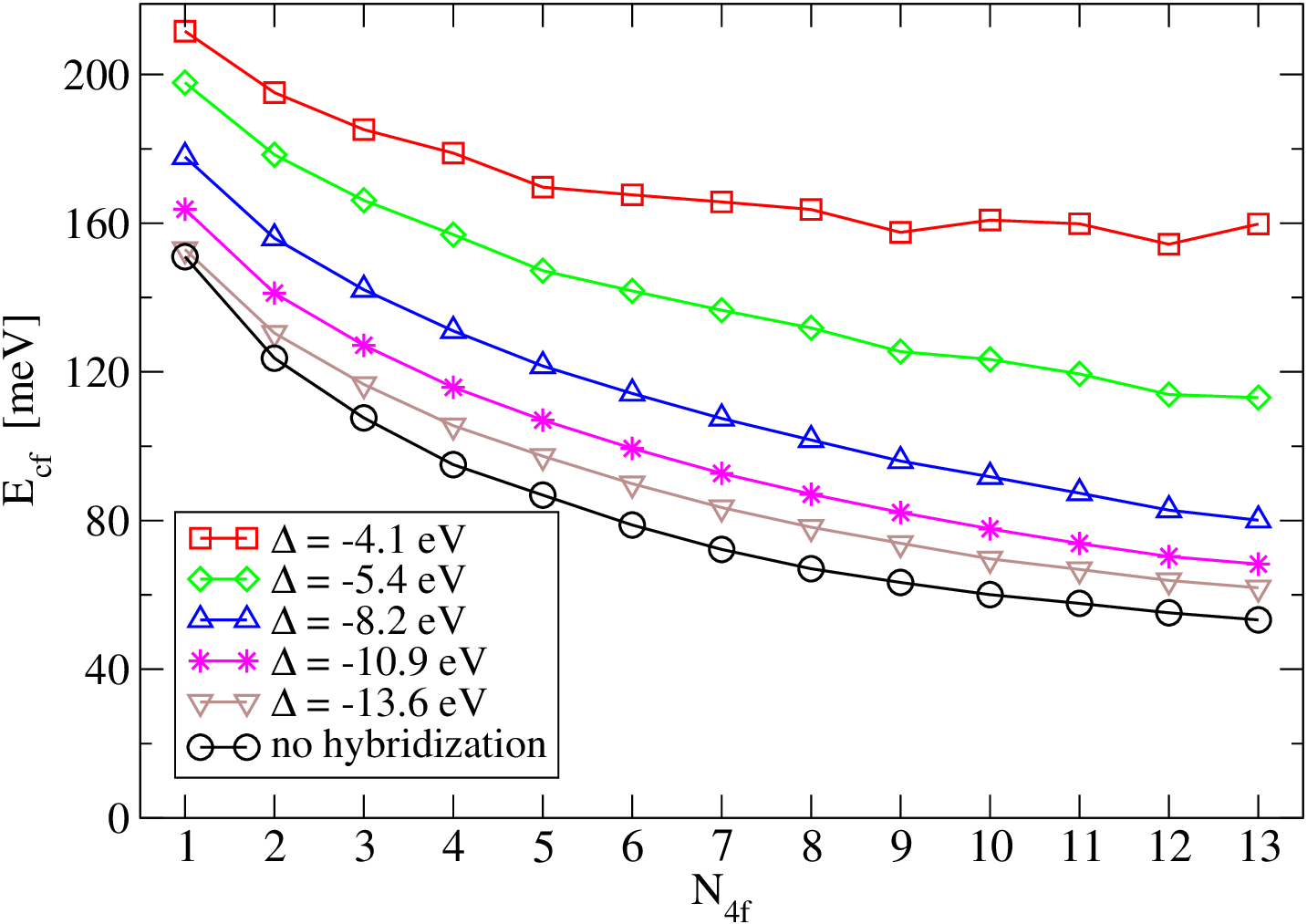}
\caption{(Color online). R:YAlO$_3$. The difference $E_{cf}$ of the lowest and highest $4f$ eigenenergy across lanthanide
series for several values $\Delta$.
The curves in this, as well as in the following figures, serve as guides for eyes only. }
\label{fig:Ecf}
\end{figure}

The density of one-particle states projected on Tb$^{3+}\;$ $4f$ orbitals, obtained from the second step
of our procedure  is shown in Fig. \ref{fig:tbdos}.
The overall strength of the crystal field may be
characterized by the difference $E_{cf}$ between the lowest and highest eigenvalues
of the crystal-field Hamiltonian $H_{cf}$ (see Fig. \ref{fig:tbdos}).
The $E_{cf}$ in TbAlO$_3$ is markedly bigger than the one in Tb:YAlO$_3$,
indicating a large effect of the local geometry - in both cases this geometry was determined by minimizing the
atomic forces. The dependence of  $E_{cf}$ on the number $N_{4f}$ of the $4f$ electrons and for several values of $\Delta$
is shown in Fig. \ref{fig:Ecf}. As expected the effect of the $f$--$p$ hybridization decreases with increasing
$4f$ - ligand level separation controlled by $\Delta$.
For fixed $\Delta$ the $E_{cf}$ exhibits a monotonous dependence on the rare earth element
characterized by $N_{4f}$.

\begin{figure}
\includegraphics[width=12cm]{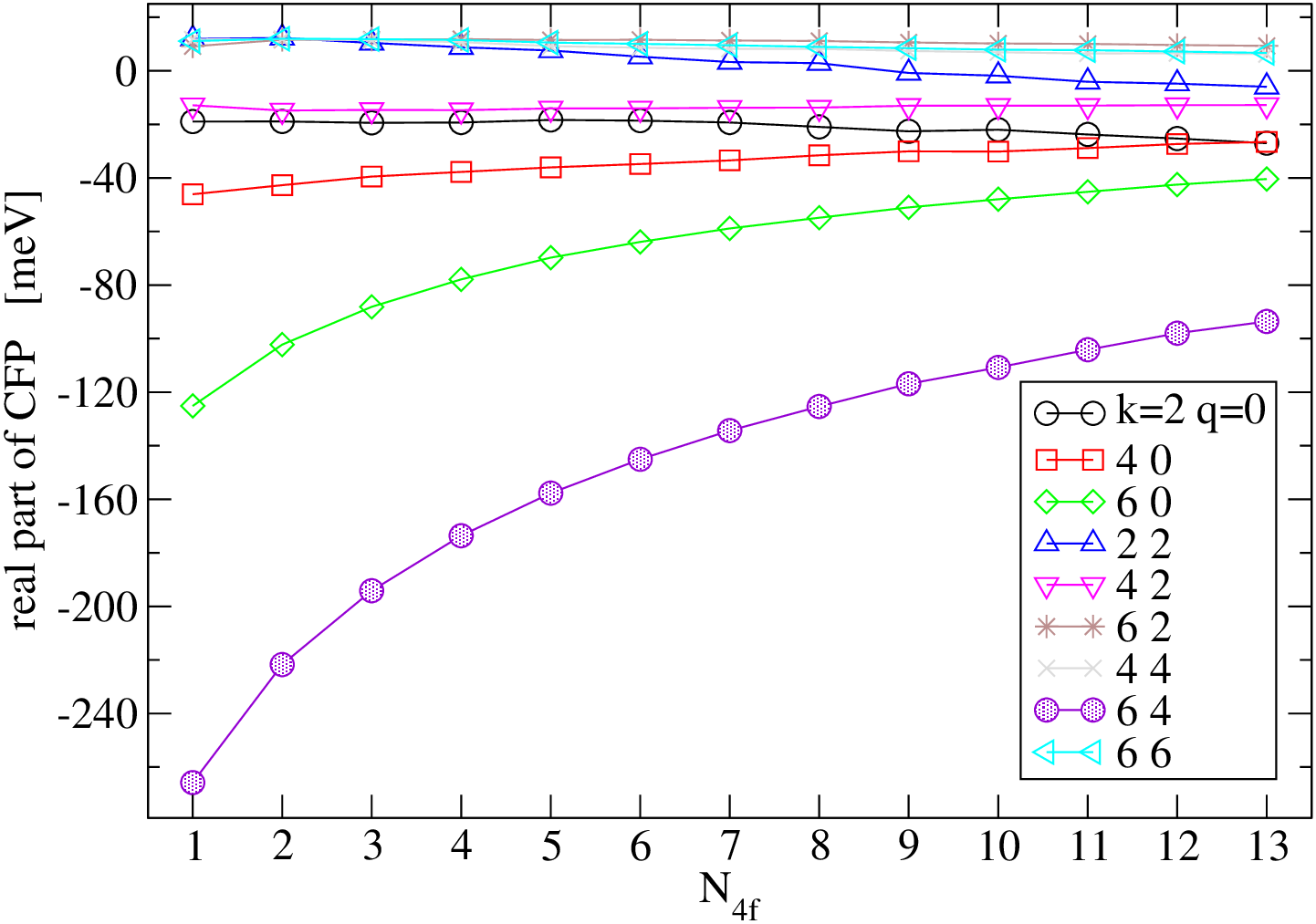}
\caption{(Color online). R:YAlO$_3$. Dependence of real part of crystal field parameters on number of the $4f$ electrons.
Shift $\Delta$= -8.2 eV.  }
\label{fig:Breal}
\end{figure}

\begin{figure}
\includegraphics[width=12cm]{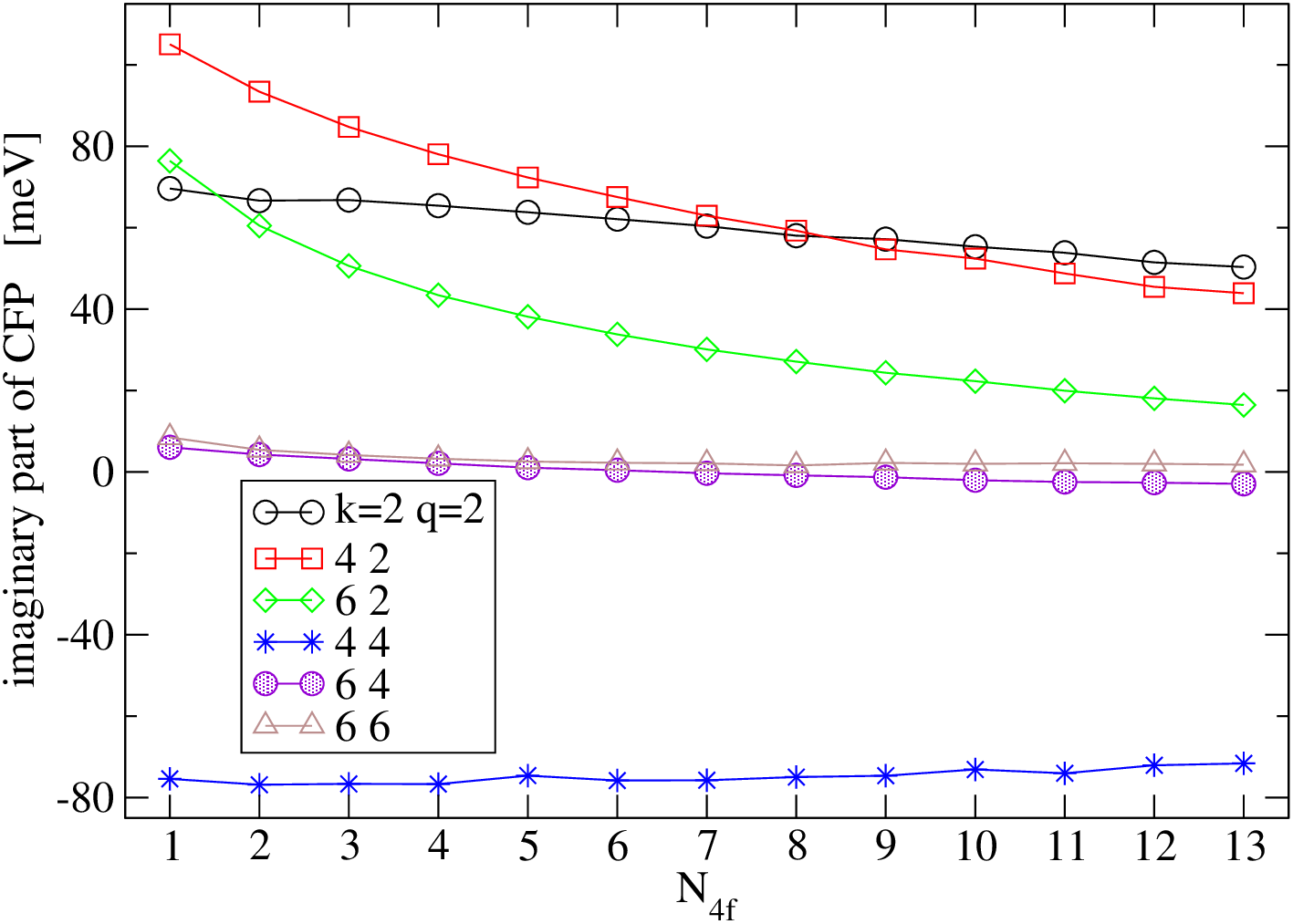}
\caption{(Color online). R:YAlO$_3$. Dependence of imaginary part of crystal field parameters on number of the $4f$ electrons.
Shift $\Delta$= -8.2 eV.  }
\label{fig:Bimag}
\end{figure}

The full information about the crystal field in the form of
the nonzero $B_{q}^{(k)}$ is shown in Figs. \ref{fig:Breal} and \ref{fig:Bimag}
using typical value of $\Delta$= -8.2 eV.
The dependence of $B_{q}^{(k)}$ on $N_{4f}$ is again smooth, the largest term being $B_{4}^{(6)}$.

We remind the reader that  the $4f$(R) states were allowed to hybridize only with the oxygen $2p$ and $2s$ states.
For Er:YAlO$_3$ we have analyzed this approximation in details. The eigenvalue problem was solved with the oxygen
states shifted by $\Delta$=-8.2 eV and, in addition, the energy of a selected valence state
(valence $s, p$ or $d$ states of Er, Y and Al) was left
unshifted, allowing thus its hybridization with the $4f$(Er). The $4f$ energies were then compared with calculation in
which the hybridization was prevented. The mean change of the $4f$ energy  was in all cases smaller than   1 meV,
with the exception of Er $5d$ states, where it reached 1.7 meV.

\section{Comparison with experiment}
\label{sec:comparison}
In order to establish how well the calculated CFP
describe the actual materials
we have calculated the crystal field splitting of the $4f^n$ atomic multiplets
in TbAlO$_3$, Nd:YAlO$_3$ and Er:YAlO$_3$, for which detailed experimental
data exist.~\cite{duan,gruber}
To this end we have solved the eigenvalue problem for the
effective Hamiltonian (1) using the 'lanthanide' code.~\cite{edwardsson}
To treat the different R$^{3+}$ ions on the same footing  we used the atomic parameters ($\hat{H}_A$) of
Carnall {\it et al.}.~\cite{carnall}
Using alternative sets of the atomic parameters \cite{duan,gruber} led to
marginal changes of the results.
 \begin{figure}
\includegraphics[width=12cm]{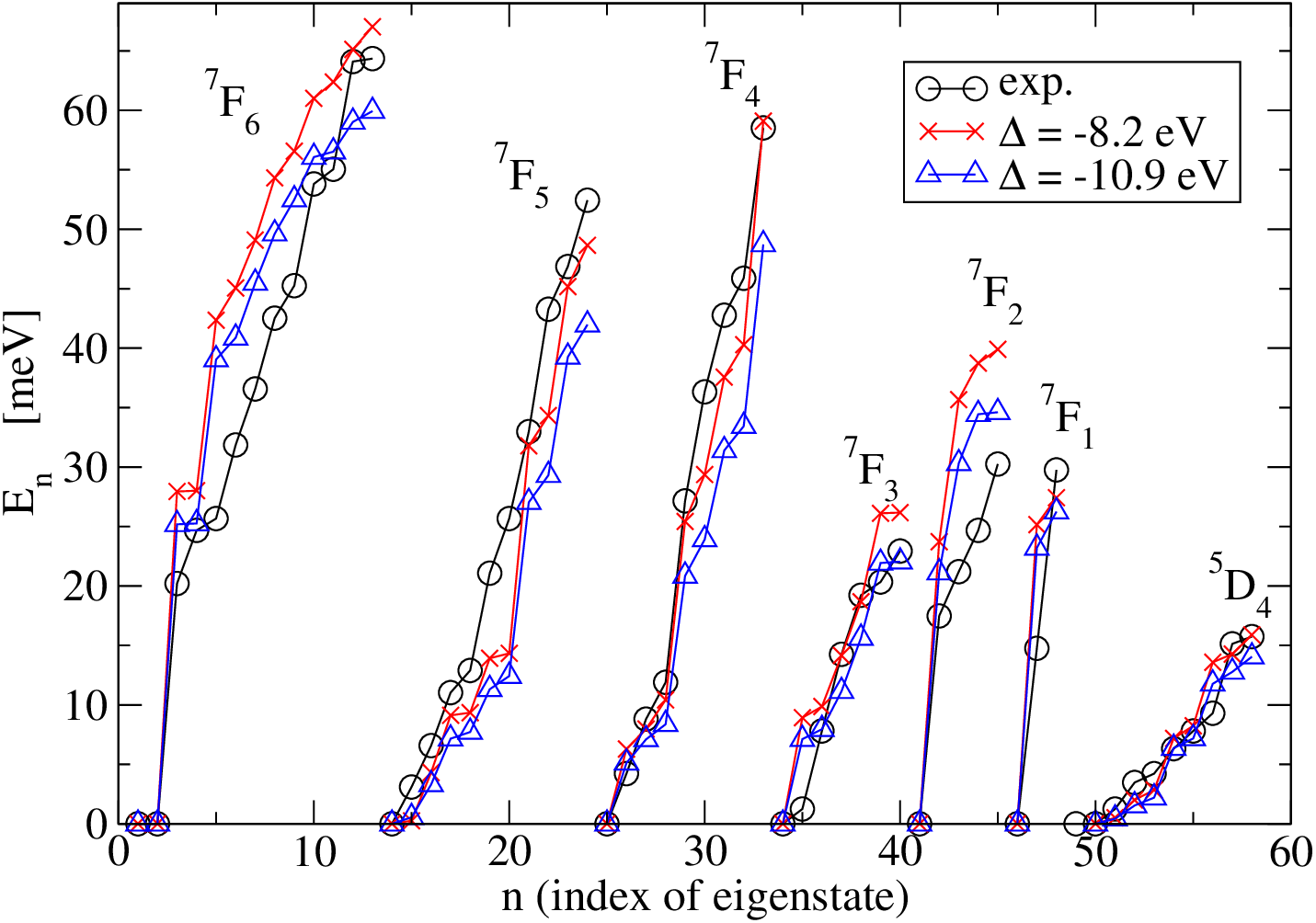}
\caption{(Color online)TbAlO$_3$. Energies of  Tb$^{3+}$ eigenstates taken relative to the lowest energy of
the multiplet. The experimental data were determined by Gruber {\it et al.} \cite{gruber}.  }
\label{fig:tbal}
\end{figure}
\begin{figure}
\includegraphics[width=12cm]{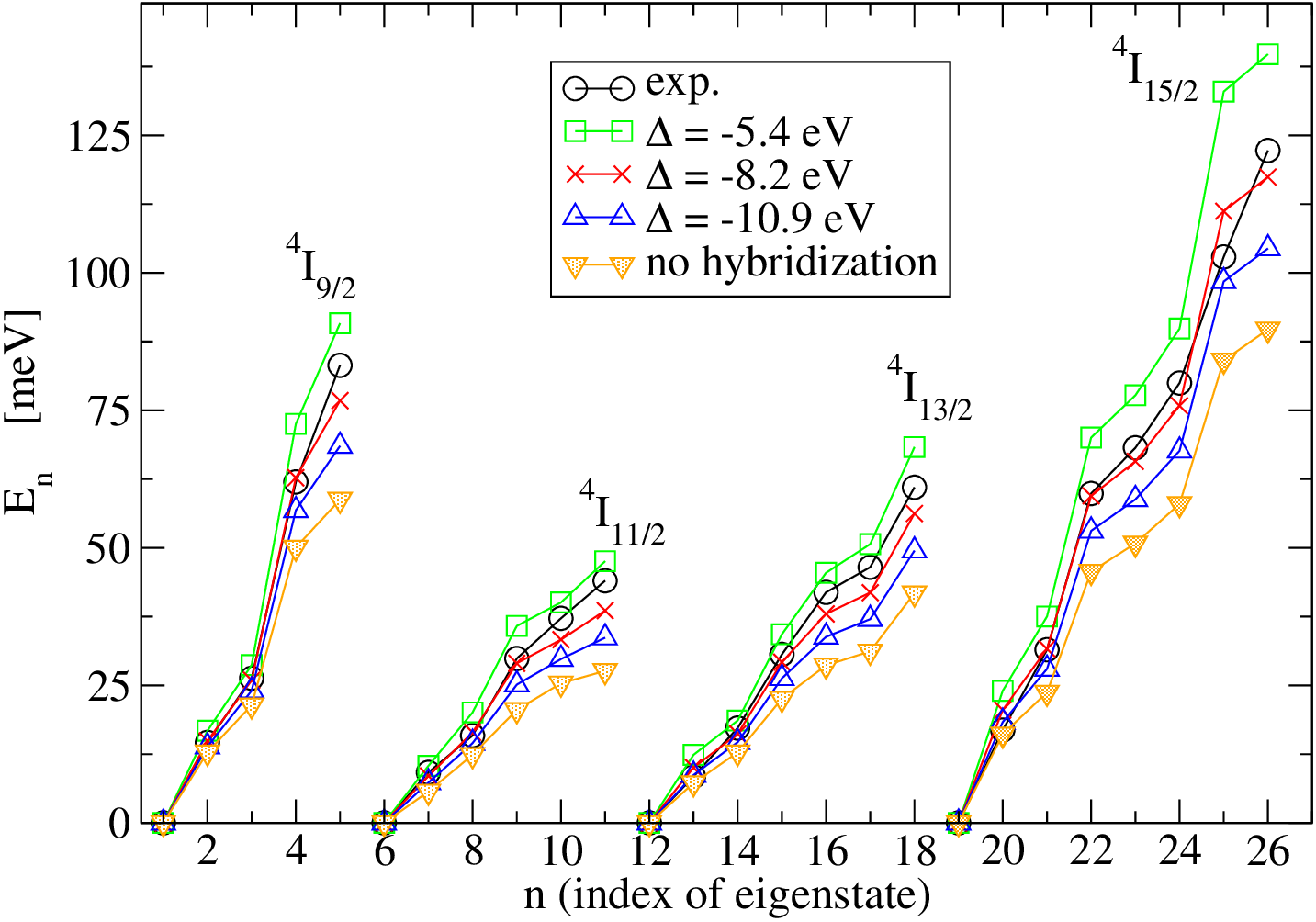}
\caption{(Color online)Nd:YAlO$_3$. Energies of  Nd$^{3+}$ eigenstates taken relative to the lowest energy of
the multiplet. The experimental data were determined by Kaminskii \cite{kaminskii} and Duan {\it et al.} \cite{duan}.  }
\label{fig:nd}
\end{figure}

In Fig. \ref{fig:tbal} the crystal field splittings of the Tb$^{3+}$ seven lowest multiplets are compared
to the experimental data.~\cite{gruber}
 \begin{figure}
\includegraphics[width=12cm]{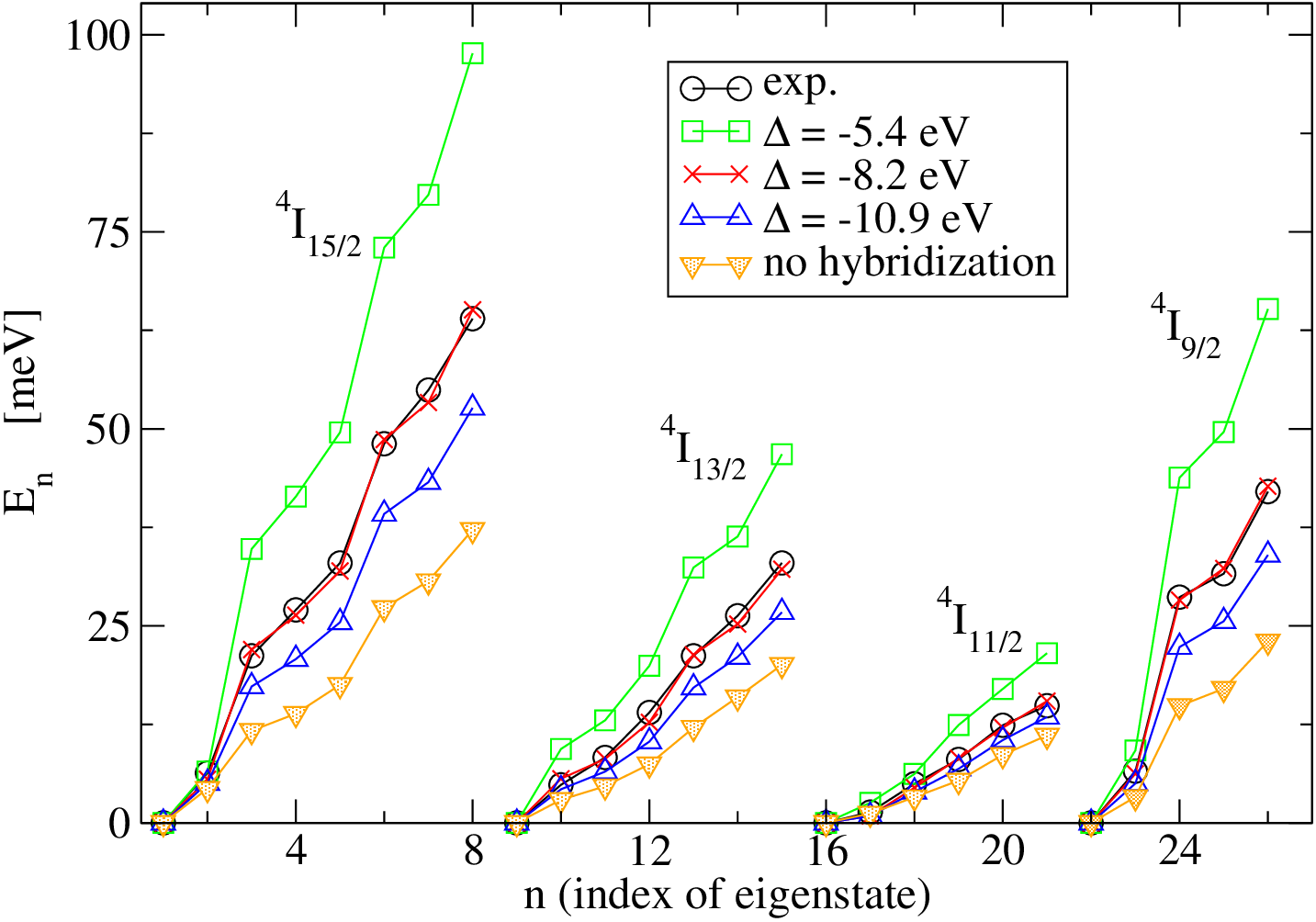}
\caption{(Color online)Er:YAlO$_3$. Energies of  Er$^{3+}$ eigenstates taken relative to the lowest energy of
the multiplet. The experimental data were determined by Donlan and Santiago \cite{donlan} and Duan {\it et al.} \cite{duan}.  }
\label{fig:er}
\end{figure}
A similar plot for the Nd$^{3+}$ in YAlO$_3$ is presented in Fig. \ref{fig:nd}. In this case the results
are more sensitive to the value of $\Delta$ and calculations for $\Delta$ = -5.4, -8.2 and -10.9 eV are confronted
with the experiment.~\cite{kaminskii,duan} In addition the results obtained with $f$--$p$ hybridization
completely eliminated are also included.
\begin{figure}
\includegraphics[width=12cm]{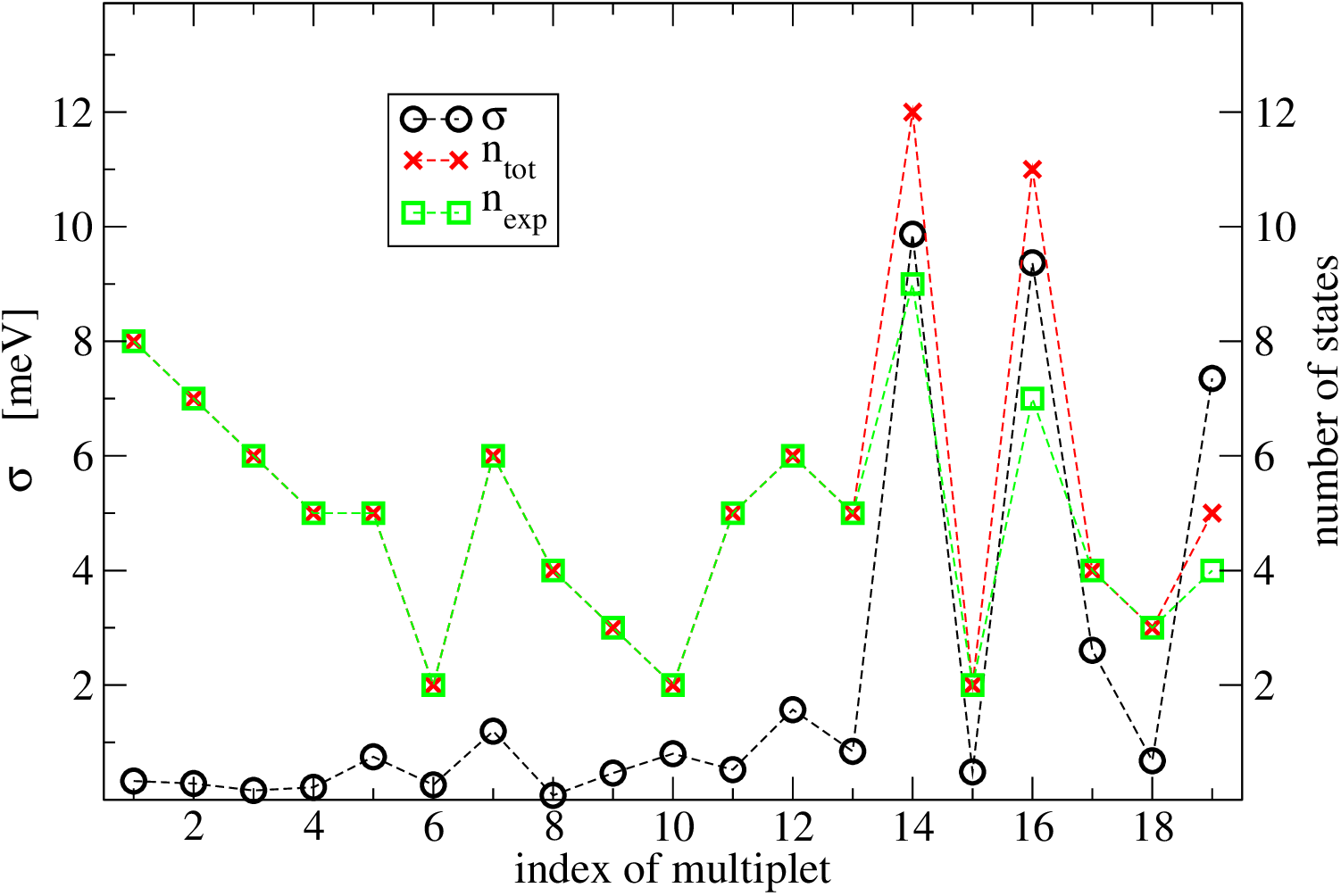}
\caption{(Color online)Er:YAlO$_3$. The mean square deviation $\sigma$ of the experimental splittings and
splittings calculated with $\Delta$= -8.2 eV (left vertical axis). The total number of levels in the set  $n_{tot}$
is denoted by crosses, number of levels observed experimentally $n_{exp}$ is denoted by squares (axis on the right).
The experimental data were determined by Donlan and Santiago \cite{donlan} and Duan {\it et al.} \cite{duan}.  }
\label{fig:sigma}
\end{figure}
As a third example we consider Er$^{3+}$ ion in YAlO$_3$. As seen in Fig. \ref{fig:er} the experimental data for the
four lowest multiplets are in a very good agreement with the calculation for $\Delta$ equal to -8.2~eV.
To quantify the agreement between the theory and all available experimental multiplet data
without overloading the reader with information we have evaluated the mean square
deviation of the experimental and calculated splittings
\begin{equation}
\sigma = \sqrt{
\frac{
\sum_{j=1}^{n_{exp}}
(E_{j,exp.} -E_{j,calc.})^2
}
{n_{exp}}
} \;,
\end{equation}
where $n_{exp}$ corresponds to the $|L,S,J \rangle$ multiplets, with the exception of the set 14, which combines
$^2K_{13/2},\; ^2P_{1/2}$ and $ ^4G_{5/2}$ multiplets. The mean square deviation in Fig.
\ref{fig:sigma} for the 19 lowest Er$^{3+}$ sets indicates a good agreement between the experiment
and theory with the exception of sets 14, 16 and 19. As shown in the figure not all levels were
observed in these three case, while for the remaining 16 multiplets the experimental information
is complete.

\section{Discussion}
\label{sec:discussion}
The results presented in Figs. 2, 6 and 7 show convincingly that the $4f$--ligand hybridization is important
amounting to about 30\% of the observed crystal-field splitting. The magnitude of our empirical estimates of $\Delta$
in the range 5-8~eV (Nd), 8-11~eV (Tb) and around 8~eV (Er) agrees with the experimentally observed trend
of the charge-transfer energies \cite{rogers} and reflects the strong Coulomb repulsion with the $4f$ shell.
\begin{figure}
\includegraphics[width=12cm]{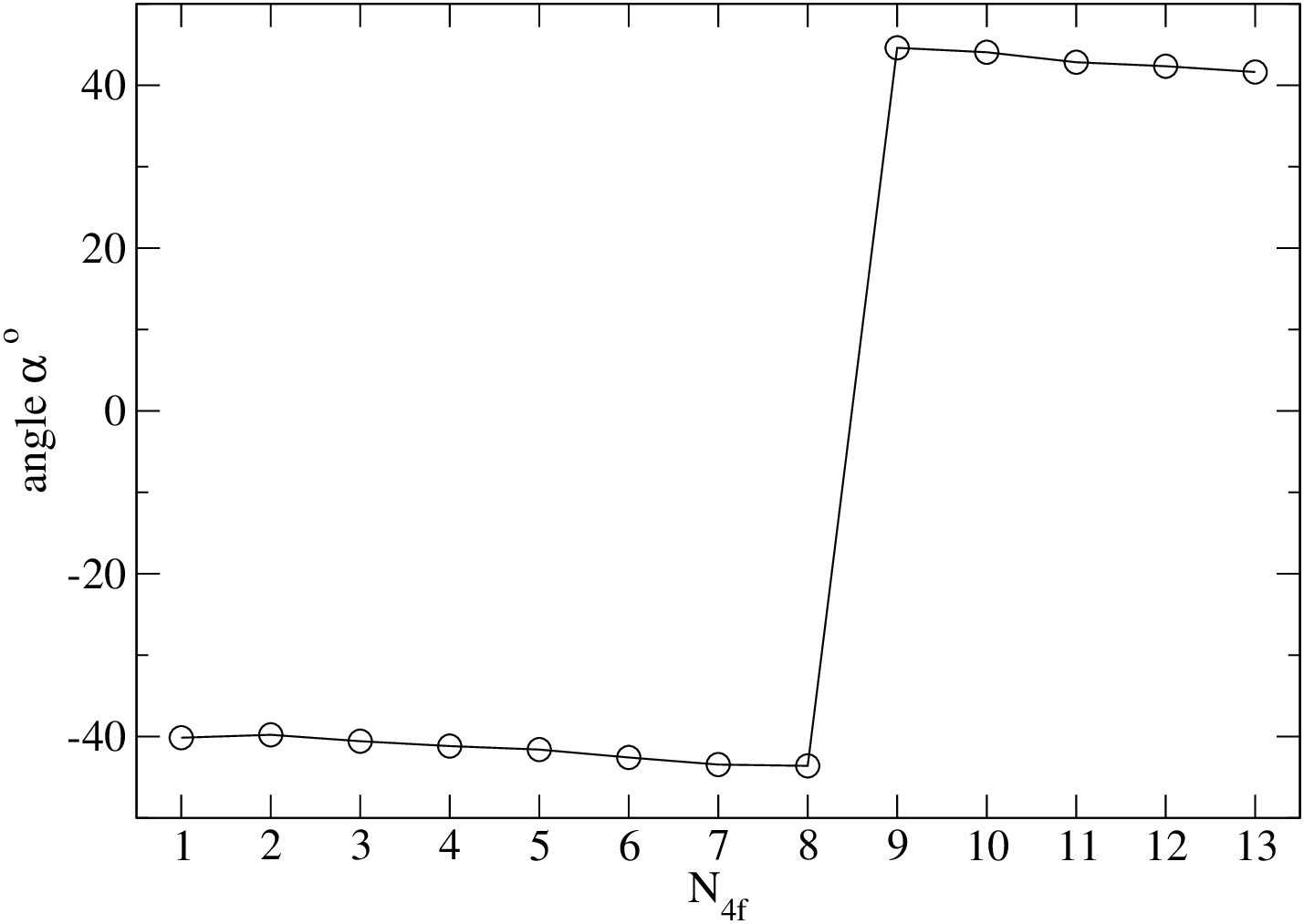}
\caption{(Color online)Angle of rotation around orthorhombic $c$ axis. }
\label{fig:angle}
\end{figure}

To compare the calculated CFP with those obtained by least squares fit to the optical spectra
\cite{duan,gruber} we have to use the coordinate system in which $B^{(2)}_2$ is real as in the
experimental analysis. This is achieved by rotation about the $c$-axis by an angle
$\alpha=-\mathrm{arctg}(\mathrm{Im}B_2^{(2)}/\mathrm{Re}B_2^{(2)})$. The value of
$\mathrm{Im}B_2^{(2)}$ is large and positive for all R$^{3+}$ ions, while $\mathrm{Re}B_2^{(2)}$
is small and positive for lighter R, changing its sign for R = Dy (see Figs.~\ref{fig:Breal} and
\ref{fig:Bimag} for $\Delta$ = -8.2~eV).
%KK V predchozi vete nebylo na prvni pohled jasne, ze podmet je ImB2, dal bych spis odkaz na Figs na konec do zavorek
The resulting $\alpha$~vs~$N_{4f}$ is shown in Fig.~\ref{fig:angle}. The $\alpha(N_{4f})$ changes
the sign, which leads to a discontinuity of the rotated $B_{q}^{(k)}(N_{4f})$ dependence despite
the fact that in a fixed coordinate system the CFP change continuously with $N_{4f}$. This
involves all the terms with $q$=2 and $q$=4. It is thus more informative to compare the absolute
values of $B_{q}^{(k)}$ than their real and imaginary parts separately. Such a comparison for
Nd$^{3+}$ and Er$^{3+}$ ions in YAlO$_3$ and for Tb$^{3+}$ ion in TbAlO$_3$ is presented in Table
\ref{tab:cfp}.
The largest contribution comes from the $B^{(6)}_4$ term and the crystal field decreases with an increasing
number of $4f$ electrons.
The CFP determined from the optical spectra suffer an ambiguity connected with numerous local minima
of the minimization in the 14 dimensional space. Returning to the Er:YAlO$_3$ case
(Fig. \ref{fig:sigma}), in which the agreement between our calculation and the incomplete experimental multiplet splittings was
significantly worse for the level sets 14, 16 and 19, we point out that assignment of the measured transitions
is far from being unique. The present calculation may serve as a useful starting point for fitting of the optical spectra.

 \begin{table}
\centering
\setcellgapes{2.5pt}
\makegapedcells
\caption{Comparison of absolute values of crystal field parameters calculated with the shift $\Delta$=-8.2 eV with those obtained by
Duan {\it et al.} \cite{duan} (exp.$^a$) and Gruber {\it et al.} \cite{gruber} (exp.$^b$) by least squares fit to optical spectra. All CFP are in units of cm$^{-1}$.}
\begin{tabular}{cccccccc}
\hline
\hline
& &\multicolumn{2}{c}{Nd:YAlO$_3$} &\multicolumn{2}{c}{Tb in TbAlO$_3$} & \multicolumn{2}{c}{Er:YAlO$_3$} \\
 %k    &  q  &  calc.  & exp. \cite{duan}  & calc.  & exp. \cite{gruber}  & calc.  & exp. \cite{duan}   \\
 \, k \,   & \, q \, & \;  calc. \;  & exp.$^a$& \;  calc. \; & exp.$^b$& \; calc. \; & exp.$^a$\\
\hline
  2 &   0 &   157 &    154  & 355 &  757 & 192 &  178    \\
  4 &   0 &   319 &    541  & 114 &  469 &  233 & 134    \\
  6 &   0 &   711 &    671  & 621 &  503 &  364 & 453    \\
  2 &   2 &   545 &  578 &  546 &  262 &  436 & 490   \\
  4 &   2 &    694 &  967 &  544 & 181 &  407 & 499   \\
  6 &   2 &    419 &  512 &  256 & 476 &  180&  208   \\
  4 &   4 &    625 &  682 &  696 &  375 & 599&  627\\
  6 &   4 &   1566 \, & 1611 \, & 1096 \, & 1235 \, &  840 &  808    \\
  6 &   6 &    101 &  132 &  210 & 512 & \; 65 & \;  78    \\
 \hline
\hline
\end{tabular}
\label{tab:cfp}
\end{table}

There are several limitations of the present method, which can be overcome
with more or less expensive methods.
Large multiplet separations, in particular near half filling of the $4f$ shell,
may lead to a non-negligible multiplet dependence of the $4f$--ligand charge-transfer energies.
It would lead to different hybridization contributions to CFP for different multiplets,
an effect considered by the correlation crystal field method \cite{judd,denning}.
The present approach can deal with such a situation by using $\Delta$ corrected
by multiplet splitting.
Given that different multiplets of Nd$^{3+}$, Tb$^{3+}$ and Er$^{3+}$ are well described
by a single CFP set each, the multiplet dependence of CFP plays only a minor role
in studied cases.

The restriction of the ligand states to those of the valence band is not always
possible. In some cases the $f^n\rightarrow f^{n-1}d$ charge transfer may be relevant.
The present method is readily applicable with the $\Delta$ shift applied to
the relevant states.
In the most general situation the charge-transfer energy is small and the perturbative
treatment of the $4f$--ligand hybridization is not justified. In such cases the
hybridization and $4f$ interaction has to be treated simultaneously, either in
a cluster calculation \cite{haverkort} or solving the quantum impurity problem.\cite{nio}
In both cases, the Wannier construction including the ligand states explicitly
can be used to construct the effective Hamiltonian. Nevertheless, the analysis
of such Hamiltonian is much more demanding than the atomic calculation presented here.

The present approach can be implemented with any full potential electronic structure code.
The only prerequisites are the possibility to eliminate the non-spherical part
of the $4f$ self-interaction in the self-consistent calculation
and the possibility to construct the Wannier orbitals.

\section{Conclusions}
We have presented a method to calculate the crystal-field parameters for the $4f$ shells of
rare earth atoms from density functional theory with a single adjustable parameter $\Delta$
corresponding to the $4f$--ligand charge-transfer energy,
which can be estimated from optical experiments. We were able to obtain the crystal-field
splittings of rare-earth impurities in YAlO$_3$ within a few meV accuracy.
The simplicity of the present method makes it a useful tool to compute
optical, magnetic and thermodynamical properties of rare-earth atoms away from Kondo regime.

\appendix
\subsection{Many-body picture of $f-p$ hybridization}
In order to illustrate the origin of the adjustable parameter $\Delta$ we resort to a simplified cluster
model which includes the rare earth ion surrounded by ligand atoms and treats the two-particle
repulsion explicitly. Its Hamiltonian reads
\begin{equation}
\label{eq:cluster}
\hat{H}=\sum_{i,j}h^{\text{at}}_{ij}\hat{f}^{\dagger}_i\hat{f}^{\phantom\dagger}_j
+ \frac{U}{2}\hat{N}_f(\hat{N}_f-1)+ \hat{W}_f
+\sum_{i,k}\bigl(V_{ik}\hat{f}^{\dagger}_i \hat{p}^{\phantom\dagger}_k+ V_{ki}\hat{p}^{\dagger}_k \hat{f}^{\phantom\dagger}_i\bigr)
+\sum_k \epsilon_k\hat{p}^{\dagger}_k\hat{p}^{\phantom\dagger}_k,
\end{equation}
where the operators $\hat{f}$ and $\hat{p}$ annihilate an electron in the $4f$ and the ligand orbitals, respectively. The first
term represents the on-site (electrostatic) crystal field and the spin-orbit coupling in the $4f$ shell,
the second and third term is the electron-electron interaction within the $4f$ shell split into the SU(N) symmetric $U$ term
and the rest $\hat{W}_f$, responsible for multiplet splitting.
The fourth term describes the $4f$--ligand hybridization $\hat{V}$ and
the last term describes the site energies of the ligand orbitals.
In situations studied in this work the lowest valence state $f^n$ is well separated
from the excited states $f^{n+1}\underline{L}$ obtained by a charge transfer from ligand to the rare earth. We assume
that average charge transfer energy $\Delta_{\mathrm{fp}}=E(f^{n+1}\underline{L})-E(f^n)$, determined by the isotropic $U$
part of the interaction and the average separation of the bare $4f$ and ligand levels, is large compared to its variation,
due to the multiplet splitting ($\hat{W}_f$) and the distribution of ligand levels ($\epsilon_k$), and treat $\Delta_{\mathrm{fp}}$
as a constant.
We reduce Hamiltonian (\ref{eq:cluster}) to the $f^n$ subspace. In the process the $4f$--ligand hybridization
gives rise to corrections to the crystal field Hamiltonian $h^{\mathrm{at}}$.
We restrict ourselves to the familiar second order perturbation.
The first and third terms of (\ref{eq:cluster}) are unchanged, the second term turns into a constant. The correction
due to hybridization amounts to
\begin{equation}
\label{eq:S-W}
\langle\alpha|\hat{H}_{\text{hyb}}|\beta\rangle=
-\sum_{\gamma}\frac{\langle\alpha|\hat{V}|\gamma\rangle
\langle\gamma|\hat{V}|\beta\rangle}{\Delta_{\mathrm{fp}}},
\end{equation}
where $\alpha$ and $\beta$ are Slater determinants from the $f^n$ subspace and $\gamma$ belongs to the $f^{n+1}\underline{L}$
subspace. The only non-zero elements of $H_{\text{hyb}}$ are the diagonal ones and elements between states that differ by transfer of
a single electron between two orbitals, i.e. $H_{\text{hyb}}$ is a correction to the one-particle part of the
atomic Hamiltonian. Equation (\ref{eq:S-W}) yields for diagonal elements
\begin{gather}
\label{eq:diag}
\begin{split}
\langle\alpha | \hat{H}_{\text{hyb}} | \alpha\rangle & =-\frac{1}{\Delta_{\mathrm{fp}}}\sum_{
k,\;i\ \mathrm{empty}}
V_{ik}V_{ki} \\
        & =E_0 + \frac{1}{\Delta_{\mathrm{fp}}}\sum_{
k,\;i\ \mathrm{full}}
 V_{ik}V_{ki} \\
        & =E_0 + \frac{1}{\Delta_{\mathrm{fp}}} \sum_{i}\sum_k V_{ik}V_{ki} \langle\alpha|\hat{f}^{\dagger}_i\hat{f}^{\phantom\dagger}_i|\alpha\rangle ,
\end{split}
\end{gather}
where $E_0=-\frac{1}{\Delta_{\mathrm{fp}}}\sum_{i,k}V_{ik}V_{ki}$ is a constant shift.
To evaluate the off-diagonal elements we use the Slater determinants
\begin{align*}
&|\alpha\rangle=\hat{f}^{\dagger}_j\hat{f}^{\phantom\dagger}_i|N\rangle \\
&|\beta\rangle=|N\rangle \\
&|\gamma\rangle=\hat{f}^{\dagger}_j\hat{p}^{\phantom\dagger}_k|N\rangle,
\end{align*}
where $|N\rangle$ is a state from the $f^n$ subspace.
In this basis we get
\begin{gather}
\label{eq:off}
\begin{split}
\langle\alpha | \hat{H}_{\text{hyb}} | \beta\rangle & =-\frac{1}{\Delta_{\mathrm{fp}}}\sum_{k}
V_{jk}V_{ki}
\langle N|\hat{f}^{\dagger}_i\hat{f}^{\phantom\dagger}_j\hat{p}^{\dagger}_k\hat{f}^{\phantom\dagger}_i
\hat{f}^{\dagger}_j\hat{p}^{\phantom\dagger}_k|N\rangle
\langle N|\hat{p}^{\dagger}_k\hat{f}^{\phantom\dagger}_j\hat{f}^{\dagger}_j\hat{p}^{\phantom\dagger}_k|N\rangle \\
& = -\frac{1}{\Delta_{\mathrm{fp}}}\sum_{k}V_{jk}V_{ki}
\langle N|\hat{f}^{\dagger}_i\hat{f}^{\phantom\dagger}_j\hat{f}^{\phantom\dagger}_i\hat{f}_j^{\dagger}|N\rangle\langle N|N\rangle \\
& = \frac{1}{\Delta_{\mathrm{fp}}}\sum_{k} V_{jk}V_{ki} =
\frac{1}{\Delta_{\mathrm{fp}}}\sum_{k} V_{jk}V_{ki}\langle\alpha|\hat{f}_j^{\dagger}\hat{f}^{\phantom\dagger}_i|\beta\rangle.
\end{split}
\end{gather}
Equations (\ref{eq:diag}) and (\ref{eq:off}) describe the correction to the crystal-field Hamiltonian arising from the hybridization
to ligands. Including this correction and omitting irrelevant constants the effective Hamiltonian for the $f^n$ subspace
reads
\begin{equation}
\hat{H}=\sum_{i,j}
\bigl(
h^{\text{at}}_{ij}+\sum_k\frac{V_{ik}V_{kj}}{\Delta_{\mathrm{fp}}}
\bigr)\hat{f}^{\dagger}_i\hat{f}^{\phantom\dagger}_j+\hat{W}_f.
\end{equation}
Note that identical form of the correction would be obtained in a non-interacting system with $\Delta_{\mathrm{fp}}$ playing
the role of the difference of one-particle levels $\epsilon_f-\epsilon_p$. The shift $\Delta$ of the $p$ levels
introduced in our computational procedure thus mimics the effect of the electron-electron repulsion
within the $4f$ shell
\begin{equation}
 \Delta = (\epsilon_f - \epsilon_p) - \Delta_{fp}.
\end{equation}
The physical meaning of the energy separation of the $f$ and $p$ states after the shift is the
charge-transfer energy, i.e. energy cost of moving and electron from ligand to the rare earth ion, in the
real interacting system. The stability of the $f^n$ rare earth valence state implies that $\Delta_{\mathrm{fp}}$ is always
positive. Because of the interacting nature of the system $\Delta_{\mathrm{fp}}$ is not directly related to the positions
of $4f$ and O $p$ peaks in photoemission experiment, which may even have reversed order.
Similar considerations apply to excursions to $f^{n-1}d$ states, which have, however, a much smaller amplitude.

\begin{acknowledgments}
The work was supported by Project No.~204/11/0713 of the Grant Agency of the Czech Republic
and by the Deutsche Forschungsgemeinschaft through FOR1346.
\end{acknowledgments}

\end{document}